\documentclass{ws-procs9x6}

\begin{document}

\title{Tadpoles and Symmetries in Higgs-Gauge Unification
Theories~\footnote{\uppercase{B}ased on talks given by
\uppercase{M}.\uppercase{Q}. at \uppercase{S}tring
\uppercase{P}henomenology 2004, \uppercase{U}niversity of
\uppercase{M}ichigan, \uppercase{A}nn \uppercase{A}rbor,
\uppercase{A}ugust 1-6, 2004 and 10th \uppercase{I}nternational
\uppercase{S}ymposium on \uppercase{P}articles, \uppercase{S}trings
and \uppercase{C}osmology (\uppercase{PASCOS}'04 and \uppercase{N}ath
\uppercase{F}est), \uppercase{N}ortheastern \uppercase{U}niversity,
\uppercase{B}oston, \uppercase{A}ugust 16-22, 2004.}}

\author{C. BIGGIO}

\address{Institut de F\'\i sica d'Altes Energies (IFAE), \\
Universitat Aut\`onoma de Barcelona, \\ 
E-08193 Bellaterra (Barcelona), SPAIN\\ 
E-mail: biggio@ifae.es}

\author{M. QUIR\'OS}

\address{Instituci\'o Catalana de Recerca i Estudis Avan\c{c}ats (ICREA) and \\
Institut de F\'\i sica d'Altes Energies (IFAE), \\
Universitat Aut\`onoma de Barcelona, \\ 
E-08193 Bellaterra (Barcelona), SPAIN\\ 
E-mail: quiros@ifae.es}

\maketitle

\abstracts{In theories with extra dimensions the Standard Model Higgs
fields can be identified with internal components of bulk gauge fields
(Higgs-gauge unification). The bulk gauge symmetry protects the Higgs
mass from quadratic divergences, but at the fixed points localized
tadpoles can be radiatively generated if $U(1)$ subgroups are
conserved, making the Higgs mass UV sensitive. We show that a global
symmetry, remnant of the internal rotation group after orbifold
projection, can prevent the generation of such tadpoles. In particular
we consider the classes of orbifold compactifications $T^d/\mathbb
Z_N$ ($d$ even, $N>2$) and $T^d/\mathbb Z_2$ (arbitrary $d$) and show
that in the first case tadpoles are always allowed, while in the
second they can appear only for $d=2$ (six dimensions).}

\section{Introduction}

Among the possible motivations for studying theories in extra
dimensions with Higgs-gauge
unification\cite{Randjbar-Daemi:1982hi}$^\textrm{-}$\cite{Scrucca:2003ut}
there is the so called little hierarchy
problem\cite{Giudice:2003nc}. The latter consists in the one order of
magnitude discrepancy between the upper bound for the Standard Model
cutoff $\Lambda_{EW}$ coming from the requirement of stability of the
Higgs mass under radiative corrections and the lower bound arising
from the non-observation of dimension-six four-fermion
operators\cite{PDG}.

Up to now the best solution to the little (and grand) hierarchy
problem is supersymmetry. Indeed in this framework the Standard Model
cutoff is identified with the mass of supersymmetric particles, while
$R$-parity conservation induces a suppression in the loop corrections
to four-fermion operators which solves the little hierarchy problem.
However, since the minimal supersymmetric Standard Model extension is
becoming very constrained, it is useful to propose possible
alternative solutions which may fill the gap between the sub-TeV scale
required for the stability of the electroweak symmetry breaking and
the multi-TeV scale required by precision tests of the Standard Model.

One possible alternative solution is Higgs-gauge unification.  In
these theories the internal components of higher dimensional gauge
bosons play the role of the Standard Model Higgses and can acquire a
non-vanishing vacuum expectation value through the Hosotani
mechanism\cite{Hosotani:1983xw}. The Higgs mass in the bulk is
protected from quadratic divergences by the higher-dimensional gauge
theory and only finite corrections $\propto (1/R)^2$ ($R$ is the
compactification radius) can appear. The Standard Model cutoff is then
identified with $1/R$ and the little hierarchy between $1/R$ and
$\Lambda$, which is now the cutoff of the higher dimensional theory,
is protected by the higher-dimensional gauge symmetry. However at the
fixed points the bulk gauge symmetry is broken and localized terms
consistent with the residual symmetries can be generated by quantum
corrections\cite{ggh}. While a direct localized squared mass
($\sim\Lambda^2$) for the Higgs-gauge fields is forbidden by a shift
symmetry remnant of the original bulk gauge
symmetry\cite{vonGersdorff:2002us}, if a $U(1)$ symmetry is conserved
at a given fixed point then the corresponding field strength can be
radiatively generated, giving rise to a quadratic divergent mass for
the Higgs\cite{vonGersdorff:2002us,Csaki:2002ur}. This is a generic
feature of orbifold compactifications in dimensions $D\geq 6$ and has
been confirmed in six-dimensional orbifold field\cite{Scrucca:2003ut}
and ten-dimensional string\cite{GrootNibbelink:2003gb} theories. One
way out\cite{Scrucca:2003ut} is that local tadpoles vanish globally,
but a more elegant way, explored in Ref.~[\refcite{Biggio:2004kr}] and
discussed in this talk, is to find a symmetry which forbids the
generation of these localized terms. This symmetry is precisely the
subgroup of the tangent space group $SO(D-4)$ whose generators commute
with the orbifold group elements leaving the considered fixed point
invariant. How this symmetry comes out and how it is related to the
generation of localized tadpoles will be discussed in the following.

\section{Symmetries at the fixed points and allowed localized terms}

We consider a gauge theory (gauge group $\mathcal{G}$) coupled to
fermions in a $D=d+4>4$ dimensional space-time parametrized by
coordinates $x^M=(x^\mu, y^i)$ where $\mu=0,1,2,3$ and $i=1,\dots
,d$. The Lagrangian is
\begin{equation} 
{\mathcal L}_{D}=-\frac{1}{4}
{F}_{MN}^A{F}^{AMN}
+i{\overline \Psi}\Gamma_D^M D_M{\Psi}\, ,
\label{bulk-lagr}
\end{equation}
with $F^A_{MN}=\partial_M A_N^A - \partial_N A_M^A - g f^{ABC} A_M^B
A_N^C$, $D_M=\partial_M-igA_M^AT^A$ and where $\Gamma_D^M$ are the
$\Gamma$-matrices corresponding to a $D$-dimensional space-time.  The
local symmetry of (\ref{bulk-lagr}) is the invariance under the
(infinitesimal) gauge transformations
\begin{equation}
\delta_\xi A_M^A=\frac{1}{g}\partial_M\xi^A-f^{ABC}\xi^BA_M^C,\quad
\delta_\xi \Psi=i\xi^AT^A\Psi\, .
\label{gtrans}
\end{equation}

Now we compactify the extra dimensions on an orbifold. Firstly we
construct a $d$-dimensional torus $T^d$ by modding out $\mathbb R^d$
by a $d$-dimensional lattice $\Lambda^d$ and then we define the
orbifold by modding out $T^d$ by $\mathbb G$, where $\mathbb G$ is a
discrete symmetry group acting non-freely (i.e.~with fixed points) on
it\cite{orbifold}.  The orbifold group is generated by a discrete
subgroup of $SO(d)$ that acts crystallographically on the torus
lattice and by discrete shifts that belong to the torus lattice.  The
action of $k\in\mathbb G$ on the torus is $k\cdot y=P_k\, y+u$, where
$P_k$ is a discrete rotation in $SO(d)$ and $u\in \Lambda^d$; $y$ and
$k\cdot y$ are then identified on the orbifold. Since the orbifold
group is acting non-freely on the torus there are fixed points
characterized by $k\cdot y_f=y_f$. Any given fixed point $y_f$ remains
invariant under the action of a subgroup $\mathbb G_f$ of the orbifold
group.

The orbifold group acts on fields $\phi_{\mathcal R}$ transforming as
an irreducible representation ${\mathcal R}$ of the gauge group
$\mathcal G$ as
\begin{equation}
k\cdot\phi_{\mathcal R}(y)
=\lambda^k_{\mathcal R}\otimes\mathcal P^k_\sigma\,
\phi_{\mathcal R}(k^{-1}\cdot y)
\label{orbfield}
\end{equation}
where $\lambda^k_{\mathcal R}$ is acting on gauge and flavor indices
and $\mathcal P^k_\sigma$, where $\sigma$ refers to the field spin, on
Lorentz indices. In particular one finds for scalar fields $\mathcal
P_0^k=1$ and for gauge fields $\mathcal P_1^k=P_k$ for a discrete
rotation ($\mathcal P_1^k=1$ for a lattice shift), while for fermions
$\mathcal{P}^k_{\frac{1}{2}}$ can be derived requiring the invariance
of the lagrangian under the orbifold action.  On the other side
$\lambda^k_{\mathcal R}$ depends on the gauge structure and the gauge
breaking of the orbifold action.

In general the orbifold action breaks the gauge group in the bulk
$\mathcal G=\{T^A\}$ to a subgroup $\mathcal H_f=\{T^{a_f}\}$ at the
fixed point $y_f$. It can be shown that the subgroup $\mathcal H_f$
left invariant by the orbifold elements $k\in\mathbb G_f$ is defined
by the generators that commute with $\lambda^k_{\mathcal R}$,
i.e.~$[\lambda^k_{\mathcal R},T^{a_f}_{\mathcal R}]=0$. The latter
condition must be satisfied by any irreducible representation
$\mathcal R$ of $\mathcal G$.

We now consider the effective four-dimensional lagrangian. This can be
written as:
\begin{equation}
\mathcal L_4^{eff}=\int d^d y \bigl[
\mathcal L_D+\sum_{f}\delta^{(d)}(y-y_f)\,\mathcal L_f \bigr]
\label{eff-lag}
\end{equation}
where $\mathcal L_D$ is given by (\ref{bulk-lagr}) and $\mathcal L_f$
is the most general lagrangian consistent with the symmetries
localized at the fixed point $y_f$. In order to write the most general
$\mathcal L_f$ we need to know which are the symmetries present at
each fixed point. First of all the operators must be invariant under
the action of the orbifold group [$\mathbb G_f$] and the 4D Lorentz
group [$SO(1,3)$]. Then we have to consider the bulk gauge symmetry
$\mathcal G$: when applied to the orbifold fixed points $y_f$ it
reduces to the four-dimensional gauge symmetry $\mathcal
H_f=\{T^{a_f}\}$ that applies to the four-dimensional gauge fields
$A_\mu^{a_f}$ which are also invariant under the orbifold action. This
consists in the usual gauge invariance under $\mathcal
H$-transformations\footnote{From here on we will remove for simplicity
the subscript ``f'' from the gauge group and the corresponding
generators.}  $\delta_\xi
A_\mu^a=\partial_\mu\xi^a/g-f^{abc}\xi^bA_\mu^c$. However this is not
the only symmetry generated by the original gauge symmetry $\mathcal
G$. Indeed by localizing the transformations (\ref{gtrans}) at the
orbifold fixed point $y_f$ and keeping the orbifold invariant terms
one can define an infinite set of transformations (remnant of the bulk
gauge invariance) induced by derivatives of $\xi^A$ that we can call
$\mathcal K$-transformations\cite{vonGersdorff:2002us}.  Then only
$\mathcal H$ and $\mathcal K$-invariant quantities are allowed at the
orbifold fixed points.

The presence of the remnant gauge symmetry $\mathcal K$ is very
important in order to prevent the appearance of direct mass terms for
gauge fields localized at the orbifold fixed points.  Indeed if the
gauge field $A^{\hat a}_{i}$ is invariant under the orbifold action,
where $T^{\hat a}\in\mathcal G/\mathcal H$, the remnant ``shift''
symmetry $\delta_\xi A_{i}^{\hat a}=\partial_{i}\xi^{\hat a}/g-f^{\hat
a b\hat c}\xi^b A_i^{\hat c}$ prevents the corresponding zero mode
from acquiring a localized mass.

Now that we know which are the symmetries at the fixed points we can
write the most general lagrangian $\mathcal L_f$. Invariant operators
are $(F^a_{\mu\nu})^2$, which corresponds to a localized kinetic term
for $A_\mu^a$ and $F^a_{\mu\nu}\tilde{F}^a_{\mu\nu}$ which is a
localized anomaly. Moreover if for some $(i,j)$ $F^a_{ij}$ is orbifold
invariant (this is model-dependent), $(F^a_{ij})^2$ can be non-zero at
a fixed point and if also $A_i^{\hat a}$ is orbifold invariant,
$(F^{\hat a}_{i\mu})^2$ can be present too. These last two lagrangian
terms give rise, respectively, to localized quartic couplings and
localized kinetic terms for $A_i^{\hat a}$. All these operators are
dimension four, that is they renormalize logarithmically. However if
$\mathcal H$ contains a $U(1)$ factor
\begin{equation}
F_{ij}^\alpha=\partial_i A_j^\alpha-\partial_jA_i^\alpha-gf^{\alpha \hat
b \hat c}A_i^{\hat b}A_j^{\hat c}\, ,
\label{fij}
\end{equation}
where $\alpha$ is the $U(1)$ quantum number, is $\mathcal H$ and
$\mathcal K$-invariant as well as orbifold and Lorentz invariant.  If
this operator is allowed, we expect that both a tadpole for the
derivative of odd fields, $\partial_iA^a_j$, and a mass term for the
even fields, $f^{a\hat b \hat c}A_i^{\hat b} A_j^{\hat c}$, will be
generated on the brane by bulk radiative corrections. Moreover, since
these operators have dimension two, we expect that their respective
renormalizations will lead to quadratic divergences, making the theory
ultraviolet sensitive.

Apart from the case of $D=5$ where the term $F_{ij}$ does not exist,
for $D\ge 6$ it does and we expect the corresponding mass terms to be
generated on the brane by radiative corrections. This has been
confirmed by direct computation in six-dimensional orbifold
field\cite{vonGersdorff:2002us}$^\textrm{-}$\cite{Scrucca:2003ut} and
ten-dimensional string\cite{GrootNibbelink:2003gb} theories. Of course
if these divergent localized mass terms were always present,
Higgs-gauge unification theories would not be useful in order to solve
the little hierarchy problem.  One way out can be that local tadpoles
vanish globally, but this requires a strong restriction on the bulk
fermion content\cite{Scrucca:2003ut}. A more elegant and efficient
solution would be finding another symmetry forbidding the generation
of localized tadpoles: this symmetry exists and has been studied in
Ref.~[\refcite{Biggio:2004kr}].\\

When compactifing a $d$-dimensional space to a smooth Riemannian
manifold (with positive signature), a tangent space can be defined at
each point and the orthogonal transformations acting on it form the
group $SO(d)$\cite{Witten}. When an orbifold group acts on the
manifold it also breaks the internal rotation group $SO(d)$ into a
subgroup $\mathcal O_f$ at the orbifold fixed point $y_f$. Indeed here
a further compatibility condition between the orbifold action and the
internal rotations is required. In particular, if the given fixed
point $y_f$ is left invariant by the orbifold subgroup $\mathbb G_f$,
only $\mathbb G_f$-invariant operators $\Phi_{\mathcal R,\sigma}$
couple to $y_f$, i.e.
\begin{equation}
k\cdot\Phi_{\mathcal R,\sigma}(y_f)=\Phi_{\mathcal R,\sigma}(y_f)\ .
\end{equation}
Acting on $\Phi_{\mathcal R,\sigma}$ with an internal rotation we get
a transformed operator that should also be $\mathbb
G_f$-invariant. This means, using Eq.~(\ref{orbfield}), that the
subgroup $\mathcal O_f$ is spanned by the generators of $SO(d)$ that
commute with $\mathcal P^k_{\sigma}$, i.e.~they satisfy the condition
\begin{equation}
[\mathcal O_f, \mathcal P^k_{\sigma}]=0
\label{conmutador}
\end{equation}
for $k\in\mathbb G_f$ and arbitrary values of $\sigma$.  In particular
in the presence of gauge fields $A_M=(A_\mu,A_i)$ an invariant
operator can be $F_{ij}$ with $\mathcal R=Adj$ and $\sigma=2$.  The
internal components $A_i$ transform under the action of the orbifold
element $k\in \mathbb G_f$ as the discrete rotation $P_k$. At the
orbifold fixed point $y_f$ only the subgroup $\mathcal O_f\subseteq
SO(d)$ survives and the vector representation $A_i$ of $SO(d)$ breaks
into irreducible representations of $\mathcal O_f$.

We have then identified an additional symmetry that the lagrangian
$\mathcal L_f$ at the fixed point $y_f$ must conserve. Summarizing,
the invariances that we have to take into account are the following:
four-dimensional Lorentz invariance [$SO(1,3)$], invariance under the
action of the orbifold group [$\mathbb G$], usual four-dimensional
gauge invariance [$\mathcal H$], remnant of the bulk gauge invariance
[$\mathcal K$] and invariance under rotations of the tangent space
[$\mathcal O_f$].

Now we consider the tadpole term of Eq.~(\ref{fij}) in order to see if
and when it is invariant under the last discussed symmetry.  We have
just seen that the vector representation $A_i$ of $SO(d)$ breaks into
irreducible representations of the internal rotation group $\mathcal
O_f\subseteq SO(d)$. In particular if the rotation subgroup acting on
the ($i,j$)-indices is $SO(2)$ then $\epsilon^{ij}F^\alpha_{ij}$,
where $\epsilon^{ij}$ is the Levi-Civita tensor, is invariant under
$\mathcal O_f$ and so it can be radiatively generated. On the other
hand if the rotation subgroup acting on the ($i,j$)-indices is $SO(p)$
($p>2$) then the Levi-Civita tensor would be $\epsilon^{i_1 i_2\dots
i_p}$ and only invariants constructed using p-forms would be
allowed. In other words a sufficient condition for the absence of
localized tadpoles is that the smallest internal subgroup factor be
$SO(p)\ (p>2)$.

\section{Tadpoles for $T^D/\mathbb{Z}_N$ orbifolds}

To show explicitly how the above discussed symmetry arguments apply,
we consider the class of orbifolds $\mathbb G=\mathbb Z_N$ for even
$d$.  The generator $P_N$ of the orbifold group is defined by
\begin{equation}
P_N=\prod_{i=1}^{d/2} e^{2\pi i\frac{k_i}{N}J_{2i-1,2i}}
\label{pe}
\end{equation}
where $k_i$ are integer numbers ($0<k_i<N$) and $J_{2i-1,2i}$ is the
generator of a rotation with angle $2\pi\frac{k_i}{N}$ in the plane
$(y_{2i-1},y_{2i})$. All orbifold elements are defined by $P_k=P_N^k$
$(k=1,\dots,N-1)$ and satisfy the condition $P_N^N=1$.  The generator
of rotations in the $(y_{2i-1},y_{2i})$-plane can be written as
$J_{2i-1,2i}={\rm diag}(0,\dots,\sigma^2,\dots,0)$ where the Pauli
matrix $\sigma^2$ is in the $i$-th two-by-two block. Therefore the
generator $P_N$ can be written as $P_N={\rm diag}(R_1,\dots,R_{d/2})$
where the discrete rotation in the $(y_{2i-1},y_{2i})$-plane is
defined as
\begin{equation}
R_i=\left(
\begin{array}{cc}
c_i & s_i\\
-s_i & c_i
\end{array}
\right)
\end{equation}
with $c_i=\cos(2\pi k_i/N)$, $s_i=\sin(2\pi k_i/N)$.

Let $y_f$ be a fixed point that is left invariant under the orbifold
subgroup $\mathbb G_f=\mathbb Z_{N_f}$ where $N_f\leq N$.  We now
define the internal rotation group $\mathcal O_f$ as the subgroup of
$SO(d)$ that commutes with the generator of the orbifold $\mathbb
Z_{N_f}$, $P_{N_f}$ as given by Eq.~(\ref{pe}) with $N$ replaced by
$N_f$. In general, if $N_f>2$ $\mathcal O_f$ is trivially provided by
the tensor product:
\begin{equation}
\mathcal O_f=\bigotimes_{i=1}^{d/2}SO(2)_i
\label{grupo}
\end{equation}
where $SO(2)_i$ is the $SO(2)\subseteq SO(d)$ that acts on the
$(y_{2i-1},y_{2i})$-subspace. In every such subspace the metric is
$\delta_{IJ}$ and the Levi-Civita (antisymmetric) tensor
$\epsilon^{IJ}$ ($I,J=2i-1,2i$, $i=1,\dots,d/2$) such that we expect
the tadpoles appearance at the fixed points $y_f$ as
\begin{equation}
\sum_{i=1}^{d/2}\mathcal
C_i\sum_{I,J=2i-1}^{2i}\epsilon^{IJ}F^\alpha_{IJ}\;\delta^{(d/2)}(y-y_f)\, .
\label{tadpolos}
\end{equation}
If $N_f=2$ then the generator of the orbifold subgroup $\mathbb
G_f=\mathbb Z_2$ is the inversion $P=-\mathbf{1}$ that obviously commutes
with all generators of $SO(d)$ and $\mathcal O_f=SO(d)$. In this case
the Levi-Civita tensor is $\epsilon^{i_1\dots i_d}$ and only a d-form
can be generated linearly in the localized lagrangian. Therefore
tadpoles are only expected in the case of $d=2$ $(D=6)$.

The last comments also apply to the case of $\mathbb Z_2$ orbifolds of
arbitrary dimensions (even or odd) since in that case the orbifold
generator is always $P=-\mathbf{1}$ and the internal rotation group
that commutes with $P$ is $\mathcal O_f=SO(d)$ for all the fixed
points. Again tadpoles are only expected for $D=6$ dimensions while
they should not appear for $D>6$.

Since every operator in $\mathcal L_f$ that is consistent with all the
symmetries should be radiatively generated by loop effects from matter
in the bulk (unless it is protected by some other $-$accidental$-$
symmetry) explicit calculations of the tadpole in orbifold gauge
theories should confirm the appearance (or absence) of them in
agreement with the above symmetry arguments. 

In Ref.~[\refcite{Biggio:2004kr}] we have considered the class of
orbifold compactifications $T^d/\mathbb Z_2$ and we have explicitly
calculated the tadpole at one- and two-loop level. The result we
found, in agreement with our general conclusions, is that the tadpole
is zero for $D>6$, while it can be non-zero only for $D=6$. In
particular we found that gauge and ghost contributions are always
zero, even in six dimensions, due to an extra parity symmetry that
inverts separately the internal coordinates. As for the fermion
contribution, it is always zero except for $D=6$, where it is
proportional to $\epsilon^{ij}$, in agreement with
Eq.~(\ref{tadpolos}). Since in six dimensions there are two
possibilities for $\mathcal{P}_{\frac{1}{2}}$, we observed that in one
case the result is chiral-independent, while in the other the sign
depends on the six dimensional chirality of fermions. In the latter
case this means that starting with Dirac fermions can imply a
vanishing tadpole. Unfortunately, due to the existing relation between
the tadpole and the mixed $U(1)$-gravitational anomaly, a vanishing
tadpole corresponds to a non-vanishing anomaly. The condition for a
vanishing tadpole coincides with the one for a vanishing anomaly only
when dealing with chiral fermions of equal six-dimensional chirality,
where this reduces to
\begin{equation}
\label{nulltad}
tr\left\{\lambda_{\mathcal R}T^A_{\mathcal R}\right\}=0 \, .
\end{equation}
%

\section{Conclusions and outlook}

In orbifold field theories Standard Model Higgs fields can be
identified with the internal components $A_i$ of bulk gauge
fields. Higher-dimensional gauge invariance prevents the Higgs from
acquiring a quadratically divergent mass in the bulk while a shift
symmetry, remnant of the bulk gauge symmetry at the fixed points,
forbids it on the branes. Nevertheless, if a $U(1)$ subgroup is
conserved at the fixed points, the corresponding field strength is
gauge invariant and can be radiatively generated giving rise, through
its non-abelian part, to a quadratically divergent mass for the Higgs.
However an additional symmetry must be taken into account. Indeed,
being the Higgs fields identified with the internal components of bulk
gauge fields, they transform in the vector representation of the
tangent space rotation group $SO(d)$. This is valid in general on a
$d$-dimensional compact manifold, but at the orbifold fixed points the
orbifold projection must be taken into account. This induces a
reduction of the tangent space group to the subgroup $\mathcal O_f$
whose generators commute with the orbifold subgroup leaving the
corresponding fixed points invariant. If $\mathcal O_f = SO(2)$, then
the Levi-Civita tensor with two indices $\epsilon^{ij}$ exists and the
corresponding field strength can be generated on the brane as
$\epsilon^{ij} F_{ij}$. On the contrary if $\mathcal O_f\supseteq
SO(p)$ with $p>2$ then only invariant terms built with the $p$-form
$\epsilon^{i_1\dots i_p}$ are allowed and so the tadpole cannot be
generated.  We have shown that in the class of orbifolds $T^d/\mathbb
Z_N$ ($d$ even, $N>2$) $\mathcal O_f= SO(2)\otimes SO(2)\otimes \dots
\otimes SO(2)$, i.e. tadpoles are always allowed. On the contrary on
orbifolds $T^d/\mathbb Z_2$ (arbitrary $d$) $\mathcal O_f= SO(d)$ that
means that tadpoles can be generated only in six dimensions ($d=2$).

However the absence of tadpoles seems to be a necessary but not
sufficient condition in order to build a realistic theory of
electroweak symmetry breaking without supersymmetry. First of all it
seems compulsory considering models with more than six dimensions,
since in five dimensions the absence of quartic coupling leads to too
low Higgs masses, while in six dimensions electroweak symmetry
breaking is spoiled by localized tadpoles. We have seen that for
$T^d/\mathbb Z_2$ orbifolds with $d>2$ tadpoles cannot be generated,
but in general these models predict the existence of $d$ Higgs fields,
leading to non-minimal models. Of course the conditions that preclude
the existence of quadratic divergences for Higgs fields do not forbid
the radiative generation of finite $\sim (1/R)^2$ masses, that vanish
in the $R\to\infty$ limit. Some of the above Higgs fields can acquire
different masses and even not participate in the electroweak symmetry
breaking phenomenon, depending on the models. Moreover, even if we
were able to build a model with only one Standard Model Higgs field,
its mass should be in agreement with LEP bounds and precision
measurements.

Another issue that must be addressed in order to construct a realistic
model is the flavour problem. One possibility should be putting matter
fermions in the bulk and coupling them to an odd mass that localizes
them at different locations\cite{Burdman:2002se}. This seems nice
since, for instance, it can also be used to explain fermion
replica\cite{Biggio:2003kp}, but problems can arise with CP violation
and flavour changing neutral currents\cite{Delgado:1999sv}. Another
possibility is to consider localized matter fermions which can develop
Yukawa couplings through Wilson line interactions after the heavy bulk
fermions have been integrated out\cite{Csaki:2002ur,Scrucca:2003ra}.


\end{document}